\documentstyle[preprint,tighten,aps,prc,epsfig]{revtex}  
\newcommand{\beq}{\begin{equation}}  
\newcommand{\eeq}{\end{equation}}  
\newcommand{\beqa}{\begin{eqnarray}}  
\newcommand{\eeqa}{\end{eqnarray}}

\date{today}  
\begin{document}  
\begin{titlepage}  
\begin{center}  
{\large{\bf Recoil Order Chiral Corrections to Baryon Octet Axial Currents}}  
\vspace{1.2cm}  
  
Shi-Lin Zhu$^a$, S. Puglia$^a$, M. J. Ramsey-Musolf$^{a,b}$  
\vspace{0.8cm}  
  
$^a$ Department of Physics, University of Connecticut,  
Storrs, CT 06269 USA\\  
$^b$ Theory Group, Thomas Jefferson National Accelerator Facility, Newport  
News,  
VA 23606 USA  
\end{center}  
\vspace{1.0cm}  
  
\begin{abstract}  
  
We calculate chiral corrections to the octet axial currents through ${\cal  
O}(p^3)$  
using baryon chiral perturbation theory (BCPT). The relativistic BCPT  
framework allows one  
to sum an infinite series of recoil corrections at a given order in the chiral  
expansion. We also include SU(3)-breaking operators occuring at ${\cal  
O}(p^2)$ not previously  
considered. We determine the corresponding low-energy constants (LEC's)  
from hyperon  
semileptonic decay data using a variety of infrared regularization schemes.  
We find that the  
chiral expansion of the axial currents does not display the proper  
convergence behavior, regardless  
of which scheme is chosen. We explore the implications of our analysis for  
determinations of the  
strange quark contribution to the nucleon spin, $\Delta s$.

\vskip 0.5 true cm  
  
PACS Indices: 12.39.Fe, 13.30.Ce, 14.20.Jn  
  
\end{abstract}  
\vspace{2cm}  
\vfill  
\end{titlepage}  
  
\pagenumbering{arabic}  
  
\section{Introduction}  
\label{sec1}  
  
In the pseudoscalar meson sector chiral perturbation theory (CPT) provides  
a consistent and  
systematic framework for calculating physical observables. Generally, they  
can be  
expanded  order by order in powers of ${p/ \Lambda_\chi}$, where  
$\Lambda_\chi =4\pi  
F_\pi$, $F_\pi=92.4$MeV, and $p$ is the typical small momenta or mass of  
the Goldstone  
bosons. When CPT is extended to include the baryons, a difficulty arises  
due to the presence  
of the large baryon mass. One encounters terms  
like ${m_N/ \Lambda_\chi}$, which obscure the convergence of the chiral  
expansion. To overcome  
this difficulty, heavy baryon chiral perturbation theory (HBCPT) was  
introduced  
\cite{j1,j2,ijmpe}. In this non-relativistic framework, the baryon mass  
appears only in  
vertices as powers of $1/ m_N$. One thus obtains a consistent expansion in  
two small parameters,  
$p/m_N$ and $p/\Lambda_\chi$. This approach has been applied successfully  
to a wide variety of  
baryon observables.  
  
Despite its successes, HBCPT comes with its own shortcomings. For example,  
the $1/m_N$ corrections  
are unnaturally large for some observables. In some cases, one requires a  
large number of  
higher-order terms in $1/m_N$ in order to obtain proper convergence  
behavior of the chiral  
expansion. From a conceptual standpoint, it was noted in \cite{kaiser} that  
HBCPT fails to  
produce the correct the analytical structure near threshold for the nucleon  
isovector  
electromagnetic form factors. The underlying reason is the non-relativistic  
treatment of the  
baryon propagators in HBCPT.  
  
A relativistic formulation of CPT for baryons was recently proposed in Ref.  
\cite{becher}  
and applied to nucleon electromagnetic form factors in Ref. \cite{kubis}.  
This formulation,  
which we denote BCPT (baryon CPT), circumvents the problematic  
$(m_N/\Lambda_\chi)^n$ terms  
by splitting a given chiral loop integral into an infrared sensitive term,  
$I$, and an infrared  
insensitive, or \lq\lq regular" piece, $R$. The former contains all the  
non-analytic contributions  
uniquely identified with chiral loops; the latter contains the power  
dependence on baryon mass.  
Since $R$ is also analytic in quark masses, its contribution may be  
completely absorbed into the  
appropriate terms in a chiral Lagrangian. Since the corresponding  
low-energy constants (LEC's)  
are determined entirely from fits to experimental data, the  
$(m_N/\Lambda_\chi)^n$ behavior never  
appears explicitly. Moreover, by retaining the fully relativistic form of  
the baryon propagators in  
$I$, one includes all of the recoil corrections to the non-analytic  
contributions at a given order  
in the chiral expansion. This procedure, known as \lq\lq infrared  
regularization", contrasts with  
the HBCPT approach, where one must explicitly work out the recoil  
corrections order-by-order in  
$1/m_N$.  
  
The simplifications introduced by BCPT have been explored in the case of a  
few observables.  
In Ref. \cite{kubis}, for example, it was pointed out that BCPT  
improves the convergence of the chiral expansion of the nucleon  
electromagnetic form factors as  
compared to HBCPT. Moreover, since BCPT is relativistic, analytical  
behaviour of the resulting form  
factors is correct.  
  
In this work we employ BCPT to calculate the one-loop chiral corrections to  
the  
axial currents of the octet baryons. The leading order chiral corrections  
to the  
axial currents are of the form $m_s\ln m_s$ and were first calculated in  
\cite{wise}.  
For a subtraction scale of $\mu=1$ GeV the corrections calculated in  
\cite{wise} are less than $30\%$.  
However, the correction due to the wave function renormalization was  
ignored in  
\cite{wise} as pointed out in \cite{j1}, where the same problem was treated  
with HBCPT  
formalism.  When wave function renormalization  
and vertex corrections are both included, the leading one-loop  
correction is large \cite{j1}. For example, the fit values of the $SU(3)$  
couplings at tree level  
in Ref. \cite{j1} are $D=0.80\pm  
0.14, F=0.50\pm 0.12$. The one-loop chiral corrections shifted  the best  
fit to $D=0.56\pm 0.1,  
F=0.33\pm 0.06$ \cite{j1}. Later the same authors included the intermediate  
decuplet baryon states in  
the chiral loops and found significant cancellations with the octet  
contributions \cite{j2}.  
While this cancellation suggests the importance of including the decuplet  
for obtaining proper  
convergence, inclusion of the decuplet is not sufficient in the case of  
some other observables.  
In the case of octet baryon magnetic moments, for example, one must also  
include the leading  
recoil-order ($1/m_N$) corrections \cite{puglia}.  
  
In this paper, we use BCPT to explore the effect of recoil corrections on  
the convergence of the  
chiral expansion of the octet axial currents. We also include ${\cal  
O}(p^2)$ chiral symmetry breaking  
terms not included in previous analyses. In order to maintain predictive  
power, we truncate the  
expansion at ${\cal O}(p^3)$. The number of LEC's appearing at ${\cal  
O}(p^4)$ prevents one from  
carrying out a model independent analysis. We also follow Refs.  
\cite{wise,j1} and set $m_u=m_d=0$ in  
performing numerical fits, although the formulae presented below included  
results for non-vanishing  
pion mass. We find that the impact of the ${\cal O}(p^2)$ symmetry breaking  
(SB) operators is  
noticeable. More importantly, the ${\cal O}(p^3)$ contributions --  
corresponding entirely to  
loop-generated recoil corrections -- are generally larger than the ${\cal  
O}(p^2)$ terms. Thus, the  
chiral expansion of the axial currents does not converge in the manner  
expected when decuplet  
contributions are integrated out. While the significance of the ${\cal  
O}(p^2)$ loop corrections was  
first noted in Ref. \cite{j1}, our study of the expansion through ${\cal  
O}(p^3)$ makes the  
non-convergence of the series abundantly clear. Contrary to one's na\"\i ve  
hope, inclusion of  
octet-only recoil order contributions only worsens the convergence  
properties of the axial currents.  
Whether explicit inclusion of recoil order decuplet contributions remedies  
this situation remains to be  
seen.  
  
In a related issue, the definition of the infrared loop contributions  
contains a degree of ambiguity.  
While the non-analytic quark mass dependence of $I$ is unique, this  
integral may also contain terms  
analytic in $m_q$. Whether or not one retains these analytic contributions  
explicitly is a matter of  
convention. The standard practice in HBCPT is to keep only the non-analytic  
loop effects. On the  
other hand, the authors of Refs. \cite{becher,kubis} also retain analytic  
pieces of $I$. We analyze  
the axial currents using both schemes. In this case, the difference amounts  
only to the treatment of  
${\cal O}(p^2)$ analytic contributions to wavefunction renormalization. The  
corresponding impact on  
the convergence properties of the chiral expansion is small. We cannot,  
however, determine whether  
this scheme insensitivity persists to higher order. While the integrals $I$  
for the axial currents  
contain a variety of ${\cal O}(p^4)$ and higher contributions, we truncate  
at ${\cal O}(p^3)$ for  
reasons noted earlier.  
  
\section{Infrared Regularization}  
\label{sec2}  

The motivation and formalism for BCPT are explained extensively in Refs.  
\cite{becher,kubis}.  
Interested readers may consult these two references for details. The key  
feature of  
BCPT is the so-called infrared regularization procedure. Following Refs.  
\cite{becher,kubis} we  
illustrate using the one-loop baryon self energy. The ultraviolet (UV)  
divergence of the one-loop  
integral is regulated using dimensional regularization. The regulated  
integral $H$ is then separated  
into the $I$ and $R$ pieces using Feynman parameters:  
\beqa  
H &=& -i\int  \frac{d^dk}{(2\pi)^d} {1 \over AB}  
=-i \int_0^1 dz \int  \frac{d^dk}{(2\pi)^d} {1 \over [(1-z)A+zB]^2}  
\nonumber \\  
 &=&-i \biggl\{ \int_0^\infty - \int_1^\infty \biggr\} dz  
 \int  \frac{d^dk}{(2\pi)^d} {1 \over [(1-z)A+zB]^2} = I + R~,  
\eeqa  
with $A=M^2-k^2-i\epsilon$, $B=m_N^2 -(p-k)^2 -i\epsilon$ and $M, m_N$ is  
the pseudoscalar  
and nucleon mass. The region of parameter integration for the integral $I$  
contains $z=0$.  
At the origin, the denominator is proportional to $A^2$, and thus, is  
highly infrared sensitive  
(singular in the case of the self energy). As shown in Ref. \cite{becher},  
all of the non-analytic  
$m_q$-dependence uniquely associated with the loop is contained in $I$.  
For the regular part, $R$, the Feynman parameter runs from one to infinity,  
and the result is  
analytic. Consequently, its contribution can be entirely absorbed into the  
appropriate operators  
appearing in the effective Lagrangian. In addition, if we  expand $I$ in  
terms of  
$1/m_N$, we recover the HBCPT result at each order.Thus,  
the infrared sensitive part $I$ of the corresponding relativistic diagram  
is just the sum of the leading HBCPT diagram and diagrams with $1/m_N$  
insertions  
to all orders. In other words, BCPT effectively sums the $1/m_N$ series in  
HBCPT.  
  
It is important to note that the inclusion of the full tower of recoil  
corrections renders the chiral  
counting somewhat  ambiguous. Contributions involving recoil effects have  
the generic  
form  
\begin{equation}  
\label{eq:counting}  
{m_X^2\over\Lambda_\chi^2} \mu^k f(\mu)\ \ \ ,  
\end{equation}  
where $m_X$ is a pseudoscalar mass,  
$\mu=m_X/m_N\propto\sqrt{m_q}/m_N$, and $f(\mu)$ is a recoil factor. For  
the axial currents,  
$\mu^k f(\mu)$ is non-analytic in $m_q$ and, therefore, can never be  
generated by terms in the  
effective Lagrangian. Nevertheless, one may perform a Taylor expansion of  
$f(\mu)$ in powers of  
$\mu^2$ about $\mu^2=0$.  Consequently, the quantity in Eq. (\ref{eq:counting})  
contains an infinite series of contributions of successively higher orders  
in $p$. The first  
term in the series -- obtained by replacing $f(\mu)\to f(0)$ -- is purely  
of ${\cal O}(p^{k+2})$.  
In the language of HBCPT, this first term in the series constitutes its  
leading contribution in  
the $1/m_N$ expansion. In what follows, we identify the chiral order of the  
term in  
(\ref{eq:counting}) by the order of its leading term in the $1/m_N$ expansion.  
  
We emphasize that the chiral order  of the recoil term in (\ref{eq:counting})  
is unambiguous only in the heavy baryon limit. Retention of the higher  
order terms associated with  
recoil factors is the price one must pay for maintaining the analyticity  
properties of loops implied by  
relativity and crossing symmetry. These properties are lost in HBCPT. It  
does not appear possible to  
respect the full analytic structure of chiral loops and maintain the  
standard chiral counting  
procedure simultaneously. Fortunately, in the case of the axial currents,  
we find that the {\em  
numerical} impact of setting $f(\mu)\to f(0)$ is negligible. In short, it  
is sufficient to work to  
${\cal O}(1/m_N)$ in the heavy baryon expansion in order to ascertain the  
effects of recoil.  
  
It was also argued in Ref. \cite{becher} that the baryon mass $m_N$ serves as  
a ``natural'' subtraction scale in BCPT using the infrared  
regularization. We follow this convention and set this subtraction scale  
equal to $m_N$.

\section{Axial Currents}  
\label{sec3}  
In writing down the octet axial currents, we follow standard conventions  
and notations.  
The most general meson-baryon Lagrangian at lowest order is  
\begin{eqnarray}  
{\cal L}_0&=&i\hbox{ Tr}\left({\bar B} (\gamma^\mu D_\mu -m_N)B \right)+ D  
\hbox{ Tr}\left({\bar B}\gamma^\mu\gamma_5 \{A_\mu, B\}\right)  
+F\hbox{ Tr}\left({\bar B}\gamma^\mu\gamma_5 [A_\mu, B]\right)\nonumber \\  
& &+{F_\pi^2\over4}\hbox{Tr}\left((D^\mu\Sigma)^{\dag}D_\mu\Sigma\right)  
+a \hbox{ Tr} M (\Sigma +\Sigma^\dag ),  
\label{lag0}  
\end{eqnarray}  
where  
\begin{equation}  
D_\mu\ B=\partial_\mu B +[V_\mu,  B],  
\end{equation}  
\begin{equation}\nonumber  
V_\mu={1\over 2}(\xi \partial_\mu \xi^\dag +\xi^\dag \partial_\mu \xi )  
\end{equation}  
\begin{equation}\nonumber  
A_\mu={i\over 2}(\xi \partial_\mu \xi^\dag -\xi^\dag \partial_\mu \xi )  
\end{equation}  
\begin{equation}  
\xi =e^{i{\pi\over F_\pi}}, \   \   \  \Sigma=\xi^2=e^{2i{\pi\over F_\pi}},  
\end{equation}  
\begin{equation}\nonumber  
\pi={1\over \sqrt{2}}\left(\begin{array}{lll}  
{\pi^0\over \sqrt{2}}+ {\eta\over \sqrt{6}} &\pi^+&K^+\\  
\pi^-&-{\pi^0\over \sqrt{2}}+ {\eta\over \sqrt{6}}&K^0\\  
K^-&{\bar K}^0&-{2\over \sqrt{6}}\eta  
\end{array}\right)  
\end{equation}  
\begin{equation}\nonumber  
B=\left(\begin{array}{lll}  
{\Sigma^0\over \sqrt{2}}+ {\Lambda\over \sqrt{6}} &\Sigma^+&p\\  
\Sigma^-&-{\Sigma^0\over \sqrt{2}}+ {\Lambda\over \sqrt{6}}&n\\  
\Xi^-& \Xi^0&-{2\over \sqrt{6}}\Lambda  
\end{array}\right)  
\end{equation}  
\begin{equation}\nonumber  
M=\left(\begin{array}{lll}m_u&0&0\\  
0&m_d&0\\0&0&m_s  
\end{array}\right)  
\end{equation}  
  
One may obtain vector and axial vector current operators from ${\cal L}_0$  
by including  
vector and axial vector sources in the covariant derivatives. The leading  
${\cal O}(p^0)$ operator  
contains only baryon fields and the LEC's $D$ and $F$. Axial currents  
involving both baryons and  
mesons first appear at ${\cal O}(p)$. Additional purely baryonic axial  
currents appear at ${\cal  
O}(p^2)$. They arise from the SU(3) SB Lagrangian  
\begin{eqnarray}\label{count}  
{\cal L}_1&=& {m_K^2\over \Lambda_\chi^2}\{  
d_1 \hbox{ Tr}\left({\bar B}\gamma^\mu\gamma_5 \{A_\mu, \chi_+\}B\right)  
+d_2\hbox{ Tr}\left({\bar B}\gamma^\mu\gamma_5  A_\mu B  
\chi_+\right)\nonumber \\  
& &+d_3 \hbox{ Tr}\left({\bar B}\gamma^\mu\gamma_5 \chi_+ B A_\mu\right)  
+d_4 \hbox{ Tr}\left({\bar B}\gamma^\mu\gamma_5 B\{A_\mu, \chi_+\}\right)  
\},  
\label{lag1}  
\end{eqnarray}  
where  
\begin{equation}  
\chi_+={1\over 2}(\xi^+ \chi\xi^++\xi\chi^+\xi)  
\end{equation}  
\begin{equation}\nonumber  
\chi=\left(\begin{array}{lll}0&0&0\\  
0&0&0\\0&0&1  
\end{array}\right)  
\end{equation}  
The LECs $d_{1-4}$ are expected to be order of unity in our normalization.  
There are  
two other terms involving $\chi$:  
\begin{equation}  
\hbox{ Tr}\left({\bar B}\gamma^\mu\gamma_5  [A_\mu, B] \right)  
\hbox{ Tr}(\chi_+)\ \ , \ \  \hbox{ Tr}\left({\bar B}\gamma^\mu\gamma_5  
\{A_\mu, B\} \right)  
\hbox{ Tr}(\chi_+).  
\end{equation}  
These terms do not break $SU(3)$ symmetry and can be absorbed into the  
definition of $D, F$ terms in Eq. (\ref{lag0}).  
  
Using ${\cal L}_{0,1}$ one obtains the axial current:  
\begin{eqnarray}  
J^A_\mu =&  
{1\over 2} D\hbox{ Tr}\left({\bar B}\gamma_\mu\gamma_5 \{\xi T^A\xi^\dag  
+\xi^\dag  
T^A\xi, B\}\right) \nonumber \\  
&+{1\over 2} F\hbox{ Tr}\left({\bar B}\gamma_\mu\gamma_5 [\xi T^A\xi^\dag  
+\xi^\dag  
T^A\xi, B]\right)\nonumber \\  
&+{1\over 2}d_1 {m_K^2\over \Lambda_\chi^2}  
 \hbox{ Tr}\left({\bar B}\gamma^\mu\gamma_5 \{\xi T^A\xi^\dag +\xi^\dag  
T^A\xi, \chi_+\}B\right) \nonumber \\  
&+{1\over 2}d_2 {m_K^2\over \Lambda_\chi^2}  
\hbox{ Tr}\left({\bar B}\gamma^\mu\gamma_5 (\xi T^A\xi^\dag +\xi^\dag  
T^A\xi ) B  
\chi_+\right)\nonumber \\  
&+{1\over 2}d_3 {m_K^2\over \Lambda_\chi^2}  
\hbox{ Tr}\left({\bar B}\gamma^\mu\gamma_5 \chi_+ B(\xi T^A\xi^\dag  
+\xi^\dag T^A\xi  
)\right)\nonumber \\  
&+{1\over 2}d_4 {m_K^2\over \Lambda_\chi^2}  
 \hbox{ Tr}\left({\bar B}\gamma^\mu\gamma_5 B\{\xi T^A\xi^\dag +\xi^\dag  
T^A\xi, \chi_+\}\right) \nonumber \\  
 &+{1\over 2}\hbox{ Tr}\left({\bar B}\gamma_\mu [\xi T^A\xi^\dag  
-\xi^\dag T^A\xi, B]\right) \nonumber \\  
&+{i\over 2}F_\pi^2\hbox{Tr}\ T^A \left((\partial_\mu\Sigma)^{\dag}\Sigma-  
\partial_\mu\Sigma \Sigma^+\right) .  
\label{current}  
\end{eqnarray}  
  
Renormalized matrix elements of $J^A_\mu$ between octet baryon states may  
be written as  
\begin{eqnarray}\label{ren}  
\langle B_i| J^A_\mu | B_j\rangle &=&  
\{ \alpha_{ij} +{\bar \alpha}_{ij}{m_K^2\over \Lambda_\chi^2}  
+ [\lambda_{ij}^\pi I_d^\pi+\lambda_{ij}^K I_d^K +\lambda_{ij}^\eta  
I_d^\eta]\alpha_{ij}\\  
\nonumber  
&&+[\beta_{ij}^\pi I_a^\pi +\beta_{ij}^K I_a^K +\beta_{ij}^\eta I_a^\eta]  
+[\gamma_{ij}^\pi I_b^\pi +\gamma_{ij}^K I_b^K +\gamma_{ij}^\eta  
I_b^\eta]\alpha_{ij}\\  
&&+[\theta_{ij}^\pi I_c^\pi +\theta_{ij}^K I_c^K +\theta_{ij}^\eta  
I_c^\eta]\alpha_{ij}\}  
{\bar u}_{B_i}\gamma_\mu\gamma_5 u_{B_j}  
\end{eqnarray}  
where the first term on the right hand side is the lowest order one. In  
terms of the $D$ and  
$F$ coefficients, this term reads for different octet states  
\begin{eqnarray}\nonumber  
\alpha_{pn}^{1+i 2}=(D+F)\; ,&\\ \nonumber  
\alpha_{\Lambda\Sigma^-}^{1+i 2}={2\over \sqrt{6}}D\; ,&\\ \nonumber  
\alpha_{\Xi^0\Xi^-}^{1+i 2}=(D-F)\; ,&\\ \nonumber  
\alpha_{p\Lambda}^{4+i 5}=-{1\over \sqrt{6}}(D+3F)\; ,&\\ \nonumber  
\alpha_{\Lambda\Xi^-}^{4+i 5}=-{1\over \sqrt{6}}(D-3F)\; ,&\\ \nonumber  
\alpha_{n\Sigma^-}^{4+i 5}=(D-F)\; ,&\\  
\alpha_{\Sigma^0\Xi^-}^{4+i 5}={1\over \sqrt{2}}(D+F)  
=\sqrt{2}\alpha_{\Sigma^+\Xi^0}^{4+i 5}  
\end{eqnarray}  
where the superscript denotes the corresponding SU(3) indices of the current.  
  
The second term arises from the SB terms in Eq. (\ref{count}). The coefficients  
${\bar \alpha}_{ij}$ are  
\begin{eqnarray}\nonumber  
{\bar \alpha}_{pn}^{1+i 2}=d_2\; ,&\\ \nonumber  
{\bar \alpha}_{\Lambda\Sigma^-}^{1+i 2}=0\; ,&\\ \nonumber  
{\bar \alpha}_{\Xi^0\Xi^-}^{1+i 2}=d_3\; ,&\\ \nonumber  
{\bar \alpha}_{p\Lambda}^{4+i 5}=-{1\over \sqrt{6}}(d+3f+2d_2)\; ,&\\ \nonumber  
{\bar \alpha}_{\Lambda\Xi^-}^{4+i 5}=-{1\over \sqrt{6}}(d-3f+2d_3)\; ,&\\  
\nonumber  
{\bar \alpha}_{n\Sigma^-}^{4+i 5}=(d-f)\; ,&\\  
{\bar \alpha}_{\Sigma^0\Xi^-}^{4+i 5}={1\over \sqrt{2}}(d+f)  
=\sqrt{2}{\bar \alpha}_{\Sigma^+\Xi^0}^{4+i 5}  
\end{eqnarray}  
where  
\begin{equation}  
d={d_1+d_4\over 2}, \ \ \ f={d_1-d_4\over 2}.  
\end{equation}  
  
The remaining terms arise from the loops of Figs 1 and 2. The coefficients  
$\lambda_{ij}^X$,  
$\beta_{ij}^X$, $\gamma_{ij}^X$ and $\theta_{ij}^X$ are given in Tables  
\ref{tab1}-\ref{tab3} and  
Eq. (\ref{eq:seagull}) below. In presenting the loop results,  
we give complete expressions for the infrared integrals $I$. When fitting  
the LEC's $D$, $F$,  
and $d_1,\ldots,d_4$, however, we include only the pieces occuring through  
${\cal O}(p^3)$. In  
doing so,  we follow approach used in Ref. \cite{becher}  
in making the chiral expansion of the loop integrals.The denominator is always  
kept intact while we expand the numerator up to order ${\cal O}(p^3)$ only  
after finishing  
the loop integral explicitly. Meanwhile we never expand terms like  
\begin{equation}  
\left(4-m_K^2/m_N^2\right)^{\pm 1/2}  
\end{equation}  
in order to preserve the analyticity properties of the integrals.  
  
\medskip  
\noindent{\bf Wavefunction renormalization}  
\medskip  
  
The third term in Eq. (\ref{ren}) arises from the wave function  
renormalization in Fig. 1. We have  
\begin{eqnarray}\nonumber  
Z_i=1+\lambda_{ii}^\pi I_d^\pi+\lambda_{ii}^K I_d^K +\lambda_{ii}^\eta  
I_d^\eta \; ,&\\  
\sqrt{Z_iZ_j}=1+\lambda_{ij}^\pi I_d^\pi+\lambda_{ij}^K I_d^K  
+\lambda_{ij}^\eta I_d^\eta  
\end{eqnarray}  
The coefficients $\lambda_{ij}^X$ ($X=\pi, K, \eta$) are collected in Table  
\ref{tab1}. The  
function  
$I_D^X$ is defined as  
\begin{equation}  
I_d^X={1\over 4}\{ 4m_N^2 m_X^2  
J^X_A(0)-\Delta_X-2m^2_X [I^X(m_N^2)+m^2_XJ^X_A(0)]\}  
\end{equation}  
where the expressions of $J^X_A(0), \Delta_X, I^X(m_N^2)$ are:  
\beqa  
\Delta_X &=&  ({m_X\over \Lambda_\chi})^2\log\mu^2 ~~,\\  
I^X(m_N^2) &=&  \frac{\mu}{\Lambda_\chi^2}\{-\mu\log\mu  
  + \frac{\mu}{2}-\sqrt{4-\mu^2}  
  \arccos\biggl(-\frac{\mu}{2}\biggr) \} ~~,\\  
J^X_A(0) &=& \frac{1}{m_N^2\Lambda_\chi^2}\{-  
  \log\mu-\frac{1}{2}  
  +\mu\  
{\arccos\biggl(-\frac{\mu}{2}\biggr)\over \sqrt{4-\mu^2}}\}  
 ~~,  
\eeqa  
where $\mu = m_X/m_N$.  
  
\medskip  
\noindent{\bf Vertex corrections}  
\medskip  
  
The fourth term comes from the vertex correction diagram Fig. 2a. The  
coeffients  
$\beta_{ij}^X$ are collected in Table \ref{tab2}. The function $I^X_a$ is  
defined as  
\begin{eqnarray}\nonumber  
I_a^X=-\Delta_X-m_X^2 I^X(m_N^2)+m_X^4 J_A^X -2{m_X^2\over \Lambda_\chi^2}  
+{m_X^4\over m_N^2\Lambda_\chi^2}  
\end{eqnarray}  
where the last two terms arise from expanding the factors ${D-4\over D-2},  
{1\over D-2}$ around  
$D-4$ in the scalar integrals in the appendix of Ref. \cite{kubis}.  
  
The fifth term is the vertex correction from the tadpole diagram in Fig.  
2b. The  
coefficients $\gamma_{ij}^X$ are presented in Table \ref{tab3}. The function  
$I_b^X=\Delta_X$.  
  
\medskip  
\noindent{\bf Seagull graphs}  
\medskip  
  
The last term in Eq. (\ref{ren}) arises from the ${\cal O}(p)$ one-meson  
operators in  
$J^A_\mu$.  The relevant Feynman diagrams are Fig. 2c and 2d. The  
contribution from these  
diagrams is entirely of recoil order, vanishing in the $m_N\to\infty$  
limit. It was not included  
in the previous HBCPT analyses which worked to leading order in the $1/m_N$  
expansion.  
For this contribution,  we have  
\begin{eqnarray}  
I_c^X&=&-{1\over 2} m^2_X I^X(m_N^2) \\  
\label{eq:seagull}  
\theta_{ij}^X&=&-4\gamma_{ij}^X  
\end{eqnarray}  
with $X=K, \eta$.  
  
It is straightforward to verify  that we recover previous results in Refs.  
\cite{wise,j1} if we use  
the relation $m_\eta^2={4\over 3}m_K^2$ and keep only $m_K^2\ln m_K^2$  
terms in Eq. (\ref{ren}).  
  
\section{Numerical analysis and discussions}  
\label{sec4}  
  
From the expressions of $I_{a,b,c,d}^X$ we find that the full one loop result  
Eq. (\ref{ren}) contains terms of ${\cal O}(p^3)$ through ${\cal O}(p^5)$.  
The terms of odd chiral order ($p^3$, $p^5$) are entirely non-analytic,  
whereas the  
loops yield both analytic and non-analytic contributions of even chiral  
order ($p^2$,  
$p^4$). Recoil order contributions first occur at ${\cal O}(p^3)$. The  
contributions  
at this order have the form given in (\ref{eq:counting}) with $k=1$ and  
\begin{equation}  
\label{eq:recoil}  
f(\mu) = \arccos\biggl(-\frac{\mu}{2}\biggr) \times  
\left(4-\mu^2\right)^{\pm 1/2}  
\ \ \ .  
\end{equation}  
Although the $f(\mu)$ is non-analytic in the complex plane, one may  
nevertheless expand it in powers of $\mu^2$ about $\mu^2=0$ along the real  
axis. When  
multiplied by the prefactor $(m_X/\Lambda_\chi)^2\times\mu$ of  
(\ref{eq:counting}),  
the leading term in the series scales as $m_X^3/\Lambda_\chi^2 m_N$, making  
it of  
chiral order $p^3$. The remaining terms -- corresponding to successively  
higher orders in $p$ --  
represent sub-leading, one-loop recoil corrections. We note, however, that  
this series cannot  
be reproduced by any combination of operators in the effective Lagrangian.  
Each term in the  
series scales as an odd power of $m_X$, that is, as a fractional power of  
quark mass. Thus,  
the infinite series of recoil corrections given by the factors in Eq.  
(\ref{eq:recoil}) is  
unambigously associated with loops. In HBCPT, one would compute these  
sub-leading corrections  
order by order in $1/m_N$ and would be forced to truncate at some order.  
  
Loop contributions involving even powers of $p$ cannot be disentangled from  
terms  
in the effective Lagrangian. For example, both the wavefunction  
renormalization diagrams and  
the vertex corrections generate analytic contributions of ${\cal O}(p^2)$.  
Since these contributions  
are quadratic in $m_X$, one could just as well absorb them into the SB  
terms of Eq. (\ref{count}).  
Similarly, at ${\cal O}(p^4)$, one encounters a new set of SB contributions  
generated by the  
Lagrangian  
\begin{eqnarray}\label{count-1}  
{\cal L}_2&=& {m_K^4\over \Lambda_\chi^4}\{  
d_5 \hbox{ Tr}\left({\bar B}\gamma^\mu\gamma_5 \chi_+ A_\mu \chi_+ B\right)  
+d_6\hbox{ Tr}\left({\bar B}\gamma^\mu\gamma_5  A_\mu B  
\chi^2_+\right)\nonumber \\  
& &+d_7 \hbox{ Tr}\left({\bar B}\gamma^\mu\gamma_5 \chi_+ B \{A_\mu,  
\chi_+\}\right)\}  
+d_8 \hbox{ Tr}\left({\bar B}\gamma^\mu\gamma_5 A_\mu BA_\nu A^\mu\right)  
+\cdots  
\label{lag2}  
\end{eqnarray}  
where we have included only a few of the relevant terms. Since the number  
of LEC's  
appearing at this order is larger than the number of available data, we  
truncate at  
${\cal O}(p^3)$ in order to avoid introducing model assumptions.  
  
To this order, we determine the LEC's $D$, $F$ and $d_1,\ldots,d_4$ from  
hyperon semileptonic decay  
data\cite{pdg}, presented in terms of axial vector couplings in Table  
\ref{tab4}.  
Since $m_u\sim m_d << m_s$, we set $m_\pi=0$ in performing our numerical fits.  
We also follow Ref. \cite{j1} and enhance  
the errors by $0.2$ to avoid the biasing the fit to the precisely known  
$D+F$ value from neutron  
beta decay. The tree level best fit  yields $D=0.78, F=0.47$ and $F/D=0.60$  
with a $\chi^2 =0.1$  
for six data points as presented in Table \ref{tab4}.  In obtaining fits at  
${\cal O}(p^2)$ and  
beyond, we follow two different procedures for treating the analytic loop  
terms: scheme B, in  
which all the ${\cal O}(p^2)$ analytic loop contributions are kept  
explicitly, as in Refs.  
\cite{becher,kubis}; and scheme C in which these analytic terms are  
effectively absorbed into the  
${\cal O}(p^2)$ LEC's.  
  
It is interesting first to truncate at ${\cal O}(p^2)$ and determine the  
impact of the SB  
contributions. In Ref. \cite{j1}, where these terms were omitted, the best  
fit values for  
the LEC's are $D=0.56$ and $F=0.33$ with $F/D=0.6$. Inclusion of the SB  
terms shifts these  
values to $D=0.55$ and $F=0.41$ ($D=0.51$, $F=0.37$) in scheme B (C), a  
10-25\% shift.  
Similarly, in when the recoil corrections are included but SB terms  
omitted, the values for  
$D$ and $F$ are both reduced by roughly 30\% from results in Ref. \cite{j1}  
(see Tables  
\ref{tab5} and \ref{tab7}).  The full ${\cal O}(p^3)$ results yield values  
of $D$ and $F$ nearly 25\%  
smaller, with $F/D$ remaining close to $0.6$. The dominant effect arises  
from inclusion of recoil.  
  
It is also instructive to determine the numerical importance of including  
the full recoil  
factors appearing in Eq. (\ref{eq:recoil}). To that end, we make the  
replacements  
\begin{eqnarray}  
\arccos\biggl(-\frac{\mu}{2}\biggr)&\to& \frac{\pi}{2}\\  
\sqrt{4-\mu^2}&\to& 2  
\end{eqnarray}  
which amounts to retaining only the leading $1/m_N$ corrections appearing  
at ${\cal O}(p^3)$.  
Taking this limit is equivalent to working to first order in $1/m_N$ with  
HBCPT.  
In this case, the best fit values for $D$ and $F$ are essentially unchanged  
from the ${\cal O}(p^3)$  
fit for scheme C, while the SB LEC's $d_1,\ldots ,d_4$ shift somewhat.  
Since the impact of the  
SB terms relative to the recoil corrections is small, it appears that  
retention of the leading  
recoil corrections is sufficient to determine the convergence behavior of  
the expansion.  
  
More significantly, the ${\cal O}(p^3)$ contributions, arising entirely  
from recoil effects,  
are as large if not larger than the ${\cal O}(p^2)$ contributions. The  
relative importance of  
each order is shown in Tables \ref{tab6} and \ref{tab8}. To illustrate, we  
write here the results for a  
few representative channels (in scheme C):  
\begin{eqnarray}  
\label{eq:ganp}  
g^A_{np}&=&0.658[1+0.419+0.495]=1.26\\  
g^A_{p\Lambda}&=&-0.488[1-0.252+1.07]=-0.88\ \ \ ,  
\end{eqnarray}  
where the terms inside the square brackets represent the relative size of  
the order $p^0$,  
$p^2$, and $p^3$ contributions, respectively. A similar pattern holds for  
the other octet axial  
vector matrix elements. Far from improving the convergence behavior of the  
octet-only chiral expansion of the axial currents, inclusion of recoil  
corrections makes it worse.  
  
In order to explore further why the chiral corrections through ${\cal  
O}(p^2)$ are so  
significant,  we collect the values of loop integral functions  
in Eq. (\ref{ren}) in Table \ref{tab9}. First, we note that the  
contribution of the vertex  
correction from Fig. 2a is suppressed due to its small coefficients  
$\beta_{ij}^X$,  
which are cubic functions of $D, F$. Although the coefficients  
$\lambda_{ij}\alpha_{ij}$  
are also cubic in $D, F$, the coefficients of $\lambda_{ij}$ are big as can  
clearly seen in  
Table \ref{tab1}. Consequently, wavefunction renormalization has a  
significant impact.  
Moreover, the contributions from the self-energy, tadpole,  
and seagull diagrams all have the same sign as the the tree  
level axial couplings. These contributions add constructively. In addition,  
the coefficients of tadpole  
diagram and seagull diagrams are linear function of $D, F$, so they are  
enhanced relative to  
the other loops in this respect.  
  
The relative size of the recoil corrections requires further explanation.  
To illustrate, consider  
the seagull contributions. Na\"\i vely, the latter ought to be suppressed  
by roughly $m_k/m_N\sim  
1/2$ relative to the ${\cal O}(p^2)$ loop effects. However, the presence of  
the $\arccos(-\mu/2)$  
in these loops generates an additional numerical factor of $\pi$ for these  
contributions at leading  
order in the $1/m_N$ expansion.  It is both the large size of the kaon mass  
and this numerical factor  
which are responsible for the large size of the ${\cal O}(p^3)$ effects.  
Moreover, such numerical  
enhancement factors appear at higher orders as well. For example, the  
${\cal O}(p^5)$ contributions  
generated by wavefunction renormalization are also proportional to  
$\arccos(-\mu/2)$. Thus, we would  
expect the pattern shown in Eq. (\ref{eq:ganp}) to persist to higher  
orders\footnote{The problem of  
the large kaon mass in SU(3) CPT with baryons has also been addressed in  
Ref. \cite{barry}.}.  
  
Finally, we illustrate the practical consequences of axial vector  
non-convergence by  
considering the strange quark contribution to the nucleon's spin, $\Delta  
s$. As shown in  
Ref.\cite{ellis}, one may express $\Delta s$ in terms of  
the polarized structure  
function integrals  
\begin{equation}  
\Gamma_{p,n}=\int_0^1\ dx\ g_1^{p,n}(x) \ \ \  
\end{equation}  
as  
\begin{equation}  
\label{eq:dels}  
\Delta s = \frac{3}{2}[\Gamma_p+\Gamma_n]-\frac{5\sqrt{3}}{6} g^A_8  
\end{equation}  
where $g^A_8$ is the axial vector coupling associated with the matrix element  
$\langle p| J_\mu^8 |p\rangle$. The combinations of LEC's required for this  
matrix element are  
\begin{equation}  
\alpha_{pp}^8={1\over 2\sqrt{3}}(3F-D)  
\end{equation}  
\begin{equation}  
\beta_{pp}^{8,K}={1\over \sqrt{3}}({2\over 3}D^3-2D^2F)  
\end{equation}  
\begin{equation}  
\beta_{pp}^{8,\eta}={1\over 24\sqrt{3}}(3F-D)^3  
\end{equation}  
\begin{equation}  
{\bar\alpha}_{pp}^8={1\over \sqrt{3}}({1\over 2}d_2-2d_4)  
\end{equation}  
\begin{equation}  
\gamma_{pp}^{8,K}=-{3\over 2},\; \; \gamma_{pp}^{8,\eta}=0  
\end{equation}  
\begin{equation}  
\theta_{pp}^{8,K,\eta}=-4\gamma_{pp}^{8,K,\eta}  
\end{equation}  
Using our results in scheme C, we obtain  
\begin{equation}  
g_8^A = 0.11[1+0.55+1]  
\end{equation}  
where the terms in the brackets correspond to the order $p^0$, $p^2$, and  
$p^3$ contributions.  
Using this result and the world average data for the $\Gamma_{p,n}$, we obtain  
\begin{equation}  
\label{eq:delsdata}  
\Delta s = 0.14 - 0.16[1+0.55+1]  
\end{equation}  
where we have omitted the experimental error bars in the first term taken  
from the  
polarized deep inelastic scattering (DIS) data. The second term  
represents the contribution from $g_8^A$, broken by successive orders as above.  
We do not quote a total for  
$\Delta s$ given that the ${\cal O}(p^3)$ contribution from $g_8^A$ is as  
large as both  
the ${\cal O}(p^0)$ term as well as the first term on the RHS of Eq.  
(\ref{eq:delsdata}).  
Given the poor convergence behavior of the expansion of $g_8^A$, extraction  
of $\Delta s$ from  
polarized DIS data is problematic. In contrast, extractions of $\Delta s$  
from semi-inclusive  
measurements performed by the Hermes collaboration or elastic  
neutrino-nucleon scattering are not  
plagued by large SU(3)-breaking uncertainties. Whether inclusion of  
decuplet intermediate states  
reduces these SU(3)-breaking uncertainties requires further study.


\newpage  
{\bf Figure Captions}  
  
\begin{center}  
{\sf FIG 1.} {Feyman diagrams for the wave function renormalization. The  
dashed and solid line  
denotes the pseudoscalar meson and baryon respectively.}  
\end{center}  
\begin{center}  
{\sf FIG 2.} {The loop diagrams for the chiral corrections to the axial  
charge. The filled  
circle is the insertion of the axial current in Eq. (\ref{ren}). }  
\end{center}

\vspace{2cm}  
  
\begin{table}  
\begin{center}~  
\begin{tabular}{|c|c|c|c|}\hline  
 & kaon loop & $\eta$ loop&$\pi$ loop  
\\  
\hline  
$\lambda_{pn}$ & ${10\over 3}D^2-4DF+6F^2$  & ${1\over 3}D^2-2DF+3F^2$  
&$3(D+F)^2$  
           \\ \hline  
$\lambda_{\Lambda\Sigma^-}$ & ${8\over 3}D^2+8F^2$ & ${4\over 3}D^2$  
&${8\over 3}D^2+4F^2$\\  
\hline  
$\lambda_{\Xi^0\Xi^-}$ & ${10\over 3}D^2+4DF+6F^2$ & ${1\over  
3}D^2+2DF+3F^2 $&$3(D-F)^2$\\  
\hline  
$\lambda_{p\Lambda}$ & ${7\over 3}D^2-2DF+9F^2$ & ${5\over 6}D^2-DF+{3\over  
2}F^2 $&${7\over  
2}D^2+3DF+{3\over 2}F^2$\\ \hline  
$\lambda_{\Lambda\Xi^-}$ & ${7\over 3}D^2+2DF+9F^2 $ & ${5\over  
6}D^2+DF+{3\over 2}F^2 $  
              &${7\over 2}D^2-3DF+{3\over 2}F^2$ \\ \hline  
$\lambda_{n\Sigma^-}$  & ${11\over 3}D^2-2DF+5F^2 $ & ${5\over  
6}D^2-DF+{3\over 2}F^2 $  
&${13\over 6}D^2+3DF+{11\over 2}F^2$ \\ \hline  
$\lambda_{\Sigma^0\Xi^-}$  & ${11\over 3}D^2+2DF+5F^2$ & ${5\over  
6}D^2+DF+{3\over 2}F^2 $  
 &${13\over 6}D^2-3DF+{11\over 2}F^2$ \\  
\hline  
$\lambda_{pp}$  & ${10\over 3}D^2-4DF+6F^2$ & $ {1\over 3}D^2-2DF+3F^2$  
&$3(D+F)^2$\\ \hline  
$\lambda_{\Lambda\Lambda}$  & ${4\over 3}D^2+12F^2$ & ${4\over 3}D^2 $  
&$4D^2$\\ \hline  
$\lambda_{\Sigma\Sigma}$  & $4D^2+4F^2$ & ${4\over 3}D^2  $ &${4\over  
3}D^2+8F^2$\\ \hline  
$\lambda_{\Xi\Xi}$  & ${10\over 3}D^2+4DF+6F^2$ & ${1\over 3}D^2+2DF+3F^2 $  
         &$3(D-F)^2$\\ \hline  
\end{tabular}  
\end{center}  
\caption{\label{tab1} The coefficients $\lambda^X_{ij}$ for the wave function  
renormalization.}  
\end{table}

\begin{table}  
\begin{center}~  
\begin{tabular}{|c|c|c|c|}\hline  
 & kaon loop & $\eta$ loop&$\pi$ loop  
\\  
\hline  
$\beta_{pn}$ & ${-D^3+D^2F-3DF^2+3F^3\over 3}$  
& ${D^3-5D^2F+3DF^2+9 F^3\over 12}$  
&${1\over 4}(D+F)^3$ \\ \hline  
  
$\beta_{\Lambda\Sigma^-}$ & $-{1\over \sqrt{6}}D^3+{1\over \sqrt{6}}DF^2$  
& $-{\sqrt{6}\over 9}D^3$ &${2\over \sqrt{6}}D({D^2\over 3}-2F^2)$\\ \hline  
  
$\beta_{\Xi^0\Xi^-}$ & $-{D^3+D^2F+3DF^2+3F^3\over 3}$  
& ${D^3+5D^2F+3DF^2-9 F^3\over 12}$&${1\over 4}(D-F)^3$\\ \hline  
  
$\beta_{p\Lambda}$ & ${ -5D^3+15D^2F+9  
DF^2-27F^3]\over 6\sqrt{6}}$  & ${1\over \sqrt{6}}[-{1\over 6}D^3+{3\over  
2}DF^2]$  
&${\sqrt{6}\over 4}D(D^2-F^2)$ \\ \hline  
  
$\beta_{\Lambda\Xi^-}$ & $ { -5D^3-15D^2F+9  
DF^2+27F^3]\over 6\sqrt{6}}$  
 & ${1\over \sqrt{6}}[-{1\over 6}D^3+{3\over 2}DF^2] $ &${\sqrt{6}\over  
4}D(D^2-F^2)$\\ \hline  
  
$\beta_{n\Sigma^-}$  & $-{D^3+D^2F+3DF^2+3F^3\over 6} $  
& $-{1\over 6}D^3+{2\over 3}D^2F-{1\over 2}DF^2 $  
&$- {D^3-2D^2F+3DF^2+6F^3\over 6}$\\ \hline  
  
$\beta_{\Sigma^0\Xi^-}$  & ${-D^3+D^2F-3DF^2+3F^3\over 6\sqrt{2}}$  & $ -{  
D^3+4D^2F+3DF^2\over 6\sqrt{2}}$  &$- {D^3+2D^2F+3DF^2-6F^3\over 6\sqrt{2}}$  
\\  
\hline  
\end{tabular}  
\end{center}  
\caption{\label{tab2} The coefficients $\beta^X_{ij}$ for the vertex correction  
of Fig. 2a.}  
\end{table}

\begin{table}  
\begin{center}~  
\begin{tabular}{|c|c|c|c|}\hline  
 & kaon loop & $\eta$ loop&$\pi$ loop  
\\  
\hline  
$\gamma_{pn}$ & $-{1\over 2}$  & $0$  &$-1$\\ \hline  
  
$\gamma_{\Lambda\Sigma^-}$ & $-{1\over 2}$ & $0$ &$-1$\\ \hline  
  
$\gamma_{\Xi^0\Xi^-}$ & $-{1\over 2}$ & $0$&$-1$\\ \hline  
  
$\gamma_{p\Lambda}$ & $-{3\over 4}$  & $-{3\over 8}$&$-{3\over 8}$\\ \hline  
  
$\gamma_{\Lambda\Xi^-}$ & $-{3\over 4} $& $-{3\over 8}$&$-{3\over 8}$ \\ \hline  
  
$\gamma_{n\Sigma^-}$  & $ -{3\over 4}$ & $-{3\over 8}$ &$-{3\over 8}$\\ \hline  
  
$\gamma_{\Sigma^0\Xi^-}$  & $-{3\over 4}$  & $-{3\over 8} $ &$-{3\over 8}$\\  
\hline  
\end{tabular}  
\end{center}  
\caption{\label{tab3} The coefficients $\gamma^X_{ij}$ for the tadpole diagram  
Fig. 2b.}  
\end{table}

\begin{table}  
\begin{center}~  
\begin{tabular}{|c|c|c|}\hline  
 & Experimental data & tree level fit  
\\  
\hline  
$g^A_{pn}$ & $(1.2573\pm 0.0028)$  & $1.253$ \\ \hline  
  
$g^A_{\Lambda\Sigma^-}$ & $\sqrt{2\over 3}(0.742\pm 0.018)$ & $0.64$   
\\ \hline  
  
$g^A_{\Xi^0\Xi^-} $$^{\dag }$  & $-$ & $0.31$\\ \hline  
  
$g^A_{p\Lambda}$ & $-\sqrt{3\over 2}(0.718\pm 0.015)$  & $-0.90$  
\\ \hline  
  
$g^A_{\Lambda\Xi^-}$ & $\sqrt{3\over 2}(0.25\pm 0.05 )$& $0.26$   
\\ \hline  
  
$g^A_{n\Sigma^-}$  & $(0.340\pm 0.017) $ & $0.31$ \\ \hline  
  
$g^A_{\Sigma^0\Xi^-}$  & ${1\over \sqrt{2}}(1.278\pm 0.158)$  & $0.89$   
\\ \hline  
$D$  & $-$  & $0.78$ \\ \hline  
$F$  & $-$  & $0.47$ \\ \hline  
$F/D$  & $-$  & $0.60$ \\ \hline  
$\chi^2$  & $-$  & $0.1$ \\ \hline  
\end{tabular}  
\end{center}  
\caption{\label{tab4} Experimental data and ${\cal O}(p^0)$ fits  
for the axial charge from  hyperon semileptonix decays. The value in the  
bracket is the  
experimental value of $g_1/f_1$. The channel with $^\dag$ is the prediction. }  
\end{table}

\begin{table}  
\begin{center}~  
\begin{tabular}{|c|c|c|c|c|}\hline  
 & Data & Tree level fit&One loop ${\cal O}(p^2)$ Fit B&One loop ${\cal  
O}(p^3)$ Fit B  
\\  
\hline  
$g^A_{pn}$ & $(1.2573\pm 0.0028)$  & $1.253$ & $1.245$ &$1.25$\\ \hline  
  
$g^A_{\Lambda\Sigma^-}$ & $\sqrt{2\over 3}(0.742\pm 0.018)$ & $0.64$ &  
$0.60$&$0.62$\\ \hline  
  
$g^A_{\Xi^0\Xi^-} $$^{\dag }$  & $-$ & $0.31$& $0.126$&$0.29$\\ \hline  
  
$g^A_{p\Lambda}$ & $-\sqrt{3\over 2}(0.718\pm 0.015)$  & $-0.90$&  
$-0.90$&$-0.89$\\ \hline  
  
$g^A_{\Lambda\Xi^-}$ & $\sqrt{3\over 2}(0.25\pm 0.05 )$& $0.26$ &  
$0.31$&$0.30$\\ \hline  
  
$g^A_{n\Sigma^-}$  & $(0.340\pm 0.017) $ & $0.31$ & $0.35$&$0.34$\\ \hline  
  
$g^A_{\Sigma^0\Xi^-}$  & ${1\over \sqrt{2}}(1.278\pm 0.158)$  & $0.89$ &  
$0.905$&$0.90$\\ \hline  
$D$  & $-$  & $0.78$ & $0.55$&$0.41$\\ \hline  
$F$  & $-$  & $0.47$ & $0.41$&$0.26$\\ \hline  
$F/D$  & $-$  & $0.60$ & $0.75$&$0.63$\\ \hline  
$\chi^2$  & $-$  & $0.1$ & $0.01$&$0.022$\\ \hline  
$d_1$  & $-$  & $-$ & $-1.08$&$-2.75$\\ \hline  
$d_2$  & $-$  & $-$ & $0.505$&$0.88$\\ \hline  
$d_3$  & $-$  & $-$ & $-0.574$&$-0.65$\\ \hline  
$d_4$  & $-$  & $-$ & $0.82$&$-0.11$\\ \hline  
\end{tabular}  
\end{center}  
\caption{\label{tab5} Our fit with Scheme B up to ${\cal O}(p^2)$, ${\cal  
O}(p^3)$. The  
channel with $^\dag$ is the prediction. }  
\end{table}  
  
\begin{table}  
\begin{center}  
\begin{tabular}{|c|c|c|c|c|}\hline  
 & Full fit results & Tree level only& ${\cal O}(p^2) $ only& ${\cal  
O}(p^3)$ only  
\\  
\hline  
$g^A_{pn}$ & $1.25$  & $0.671$ & $0.245$ &$0.334$\\ \hline  
  
$g^A_{\Lambda\Sigma^-}$ & $0.62$ & $0.33$ & $0.079$&$0.211$\\ \hline  
  
$g^A_{\Xi^0\Xi^-} $$^{\dag }$  & $0.29$ & $0.143$& $-0.031$&$0.178$\\ \hline  
  
$g^A_{p\Lambda}$ & $-0.89$  & $-0.489$& $0.123$&$-0.524$\\ \hline  
  
$g^A_{\Lambda\Xi^-}$ & $0.30$& $0.157$ & $-0.063$&$0.206$\\ \hline  
  
$g^A_{n\Sigma^-}$  & $0.34 $ & $0.143$ & $0.053$&$0.144$\\ \hline  
  
$g^A_{\Sigma^0\Xi^-}$  & $0.90$  & $0.474$ & $-0.158$&$0.584$\\ \hline  
\end{tabular}  
\end{center}  
\caption{\label{tab6} The separation of our full up to ${\cal O}(p^3)$ fit  
results  
with Scheme B into  tree level, pure ${\cal O}(p^2)$, and ${\cal O}(p^3)$  
pieces for the sake of  
the discussion  of convergence of the chiral expansion. }  
\end{table}

\begin{table}  
\begin{center}~  
\begin{tabular}{|c|c|c|c|c|}\hline  
 & Data & Tree level fit&One loop ${\cal O}(p^2)$ Fit C&One loop ${\cal  
O}(p^3)$ Fit C  
\\  
\hline  
$g^A_{pn}$ & $(1.2573\pm 0.0028)$  & $1.253$ & $1.26$ &$1.26$\\ \hline  
  
$g^A_{\Lambda\Sigma^-}$ & $\sqrt{2\over 3}(0.742\pm 0.018)$ & $0.64$ &  
$0.64$&$0.61$\\ \hline  
  
$g^A_{\Xi^0\Xi^-} $$^{\dag }$  & $-$ & $0.31$& $0.24$&$0.22$\\ \hline  
  
$g^A_{p\Lambda}$ & $-\sqrt{3\over 2}(0.718\pm 0.015)$  & $-0.90$&  
$-0.85$&$-0.88$\\ \hline  
  
$g^A_{\Lambda\Xi^-}$ & $\sqrt{3\over 2}(0.25\pm 0.05 )$& $0.26$ &  
$0.34$&$0.30$\\ \hline  
  
$g^A_{n\Sigma^-}$  & $(0.340\pm 0.017) $ & $0.31$ & $0.35$&$0.34$\\ \hline  
  
$g^A_{\Sigma^0\Xi^-}$  & ${1\over \sqrt{2}}(1.278\pm 0.158)$  & $0.89$ &  
$0.88$&$0.90$\\ \hline  
$D$  & $-$  & $0.78$ & $0.513$&$0.39$\\ \hline  
$F$  & $-$  & $0.47$ & $0.370$&$0.26$\\ \hline  
$F/D$  & $-$  & $0.60$ & $0.72$&$0.67$\\ \hline  
$\chi^2$  & $-$  & $0.1$ & $0.004$&$0.013$\\ \hline  
$d_1$  & $-$  & $-$ & $-1.17$&$-2.73$\\ \hline  
$d_2$  & $-$  & $-$ & $0.60$&$0.82$\\ \hline  
$d_3$  & $-$  & $-$ & $-0.62$&$-0.54$\\ \hline  
$d_4$  & $-$  & $-$ & $0.84$&$0.094$\\ \hline  
\end{tabular}  
\end{center}  
\caption{\label{tab7} Our fit with Scheme C up to ${\cal O}(p^2)$, ${\cal  
O}(p^3)$. The  
channel with $^\dag$ is the prediction. }  
\end{table}  
  
\begin{table}  
\begin{center}  
\begin{tabular}{|c|c|c|c|c|}\hline  
 & Full fit results & Tree level only& ${\cal O}(p^2) $ only& ${\cal  
O}(p^3)$ only  
\\  
\hline  
$g^A_{pn}$ & $1.26$  & $0.658$ & $0.276$ &$0.326$\\ \hline  
  
$g^A_{\Lambda\Sigma^-}$ & $0.61$ & $0.318$ & $0.095$&$0.197$\\ \hline  
  
$g^A_{\Xi^0\Xi^-} $$^{\dag }$  & $0.22$ & $0.12$& $-0.042$&$0.142$\\ \hline  
  
$g^A_{p\Lambda}$ & $-0.88$  & $-0.488$& $0.136$&$-0.528$\\ \hline  
  
$g^A_{\Lambda\Xi^-}$ & $0.30$& $0.17$ & $-0.089$&$0.219$\\ \hline  
  
$g^A_{n\Sigma^-}$  & $0.34 $ & $0.12$ & $0.101$&$0.119$\\ \hline  
  
$g^A_{\Sigma^0\Xi^-}$  & $0.90$  & $0.465$ & $-0.135$&$0.57$\\ \hline  
\end{tabular}  
\end{center}  
\caption{\label{tab8} The separation of our full up to ${\cal O}(p^3)$ fit  
results  
with Scheme C into tree level, pure ${\cal O}(p^2)$, and ${\cal O}(p^3)$  
pieces.}  
\end{table}  
  
\begin{table}  
\begin{center}~  
\begin{tabular}{|c|c|c|}\hline  
 & kaon loop & $\eta$ loop  
\\  
\hline  
$I_a$ & $0.21$  & $0.27$  \\ \hline  
$I_b$ & $-0.23$  & $-0.237$  \\ \hline  
$I_c$ & $0.167$  & $0.230$  \\ \hline  
$I_d$ & $0.34$  & $0.424$  \\ \hline  
$J_A$ & $0.533$  & $0.504$  \\ \hline  
$\Delta$ & $-0.23$  & $-0.237$  \\ \hline  
$I (m_N^2) $ & $-1.37$  & $-1.53$  \\  
\hline  
\end{tabular}  
\end{center}  
\caption{\label{tab9} The values of loop integral functions in Eq.  
(\ref{ren}).}  
\end{table}

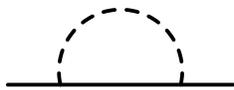
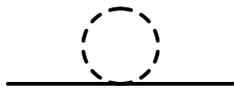
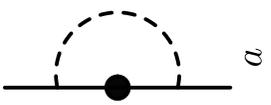
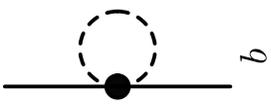
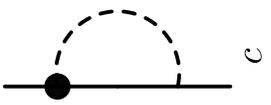
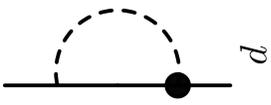
\end{document}